# Quality of Service (QoS)-driven Edge Computing and Smart Hospitals: A Vision, Architectural Elements, and Future Directions

Rajkumar Buyya[1], Satish N. Srirama[2], Redowan Mahmud[3], Mohammad Goudarzi[1], Leila Ismail[1,4], and Vassilis Kostakos[1]

[1] Cloud Computing and Distributed Systems (CLOUDS) Lab
School of Computing and Information Systems
The University of Melbourne, Australia

[2] School of Computer and Information Sciences
The University of Hyderabad, India

[3] School of Electrical Engineering, Computing and Mathematical Sciences
Curtin University, Perth, Australia

[4] Intelligent Distributed Computing and Systems (INDUCE) Lab
Department of Computer Science and Software Engineering
National Water and Energy Center
United Arab Emirates University, United Arab Emirates

**Abstract:** The Internet of Things (IoT) paradigm is drastically changing our world by making everyday objects an integral part of the Internet. This transformation is increasingly being adopted in the healthcare sector, where Smart Hospitals are now relying on IoT technologies to track staff, patients, devices, and equipment, both within a hospital and beyond. This paradigm opens the door to new innovations for creating novel types of interactions among objects, services, and people in smarter ways to enhance the quality of patient services and the efficient utilisation of resources. However, the realisation of real-time IoT applications in healthcare and, ultimately, the development of Smart Hospitals are constrained by their current Cloud-based computing environment. Edge computing emerged as a new computing model that harnesses edge-based resources alongside Clouds for real-time IoT applications. It helps to capitalise on the potential economic impact of the IoT paradigm of $11 trillion per year, with a trillion IoT devices deployed by 2025 to sense, manage and monitor the hospital systems in real-time. This vision paper proposes new algorithms and software systems to tackle important challenges in Edge computing-enabled Smart Hospitals, including how to manage and execute diverse real-time IoT applications and how to meet their diverse and strict Quality of Service (QoS) requirements in hospital settings. The vision we outline can help tackle timely challenges that hospitals increasingly face.

## 1. Introduction

Internet of Things (IoT) devices are hugely impacting our lives [1]. It is projected that the IoT paradigm will have a tremendous economic impact of $11 trillion per year, with a trillion IoT devices deployed by 2025 to monitor and manage smart systems in real-time [3] in different domains, including smart healthcare. The Internet of Medical Things (IoMT) devices, deployed for supporting applications within a smart hospital, are geographically distributed across floors, buildings, and sites and increasingly reaching into patients' own homes, especially in pandemic scenarios such as COVID-19. Collectively they can generate a tremendous amount of data in a short time. Conceptually, data in such IoT environments can be classified into Big Stream transient data (such as real-time localisation) constantly captured from IoT devices and sensors, and Big Data persistent data and knowledge (such as medical records) stored and archived in a centralised Cloud storage. Thus, applications require both big stream and big data processing capabilities for decision-making in real-time. Their unique requirements in the context of smart hospitals are summarised as follows: (a) ephemeral performance-based data (such as surgery completion times and environmental sensing) need real-time analysis to provide new approaches for governance and policy-making in hospitals; (b) surveillance and monitoring of the patient condition, both within a hospital and at a patient's home, pose strict real-time low-latency requirements with strict deadlines to enable rapid and timely response; and (c) strict privacy requirements pose additional constraints on how and where can information be transported, processed and stored, and also require a precise specification of these constraints.

These requirements impose a massive burden on the computational infrastructure that needs to cope with massive storage and processing power requirements. Although Cloud-based approaches are designed to deal with large volumes of data, they are unable to support real-time data analysis and actions [8]. With millions of nodes sensing and generating data, the current Cloud-centric IoT model of sending the data to a remote Cloud for processing is neither scalable nor meets QoS requirements of IoT-based Smart Hospital applications such as low end-to-end latency and high throughput for responding to requests or situations such as emergencies in real-time. In addition, it leads to network congestion, and the Internet needs to deal with this excess data outstripping the capacity, causing additional communication costs in accessing services.

The dynamic nature of IoT environments such as Smart Hospitals, the real-time requirements of their applications, and the increased processing capacity of edge devices have led to the emergence of the Edge computing paradigm (see Figure 1). Edge computing harnesses network-wide resources and offers lightweight Cloud-like services at the Edge of the network to support latency-sensitive IoT applications in real-time. It can integrate Cloud capabilities for scalable storage, as well as processing services on-demand, to support latency-tolerant IoT systems and applications. IoT devices are connected to Edge Gateway Nodes (EGNs), such as insulin pumps, patient tags, smartphones, tablet computers, Raspberry Pi devices, and nano servers, which are part of the end users. For each IoT application, EGNs offer interfaces for authentication, sensing frequency calibration and data aggregation. EGNs forward the sensed data to Edge Computational Nodes (ECNs), such as CISCO IOx-supported networking devices and micro data centres, offered by Edge Service Providers (ESPs), for further processing operations such as data filtration and data analysis. However, there are many challenges involved in developing and deploying IoT applications for smart hospitals. Some of these challenges are:

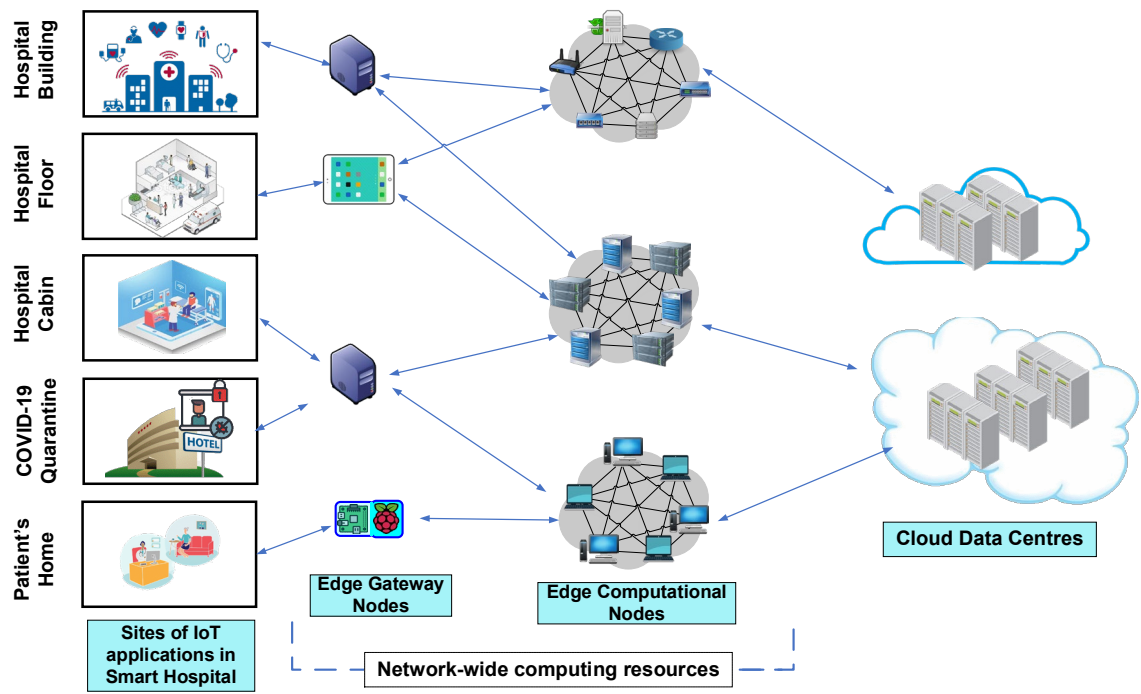

**Figure 1:** Edge computing-driven environment for IoT-applications in Smart Hospitals.

- **Quality of Services (QoS).** Unlike Clouds, Edge computing nodes are resource constrained, distributed and heterogeneous, and they are interconnected through dynamic networks [13]. Furthermore, Edge computing needs to support user mobility, interface diversity and distributed data analysis tasks to address the QoS requirements of Smart Hospitals IoT applications such as low latency service delivery.
- **Communications and Energy Efficiency.** Edge computing needs to minimise network traffic and power consumption while meeting the demands of IoT applications. Across various wards and departments of a Smart Hospital, features of IoT applications and workloads can vary significantly. IoT workloads can be either event-driven or stream-based and can encapsulate multimedia or scalar data. Applications can follow different kinds of execution models such as streams, workflows, bag of tasks, and graphs to process the IoT workloads. Runtime requirements of these applications can also change according to the nature of processing operations (e.g., compute or data-intensive). These requirements, along with the need for harnessing processing capabilities in network switches and routers to meet QoS requirements of IoT applications, make the management of resources in Edge computing-enabled Smart Hospitals a challenging task.

- **Security and Privacy.** Several features related to security need to be fulfilled, such as privacy, in which patients' sensitive data are not revealed to other users of the network, and integrity so that data is not tampered with during processing or transmission over the network. In addition, Denial of Service Attack (DoS) should be considered, as attackers' requests which look legitimate cause a decrease in IoT applications performance that could be life-threatening and an increase in energy consumption.

To address these challenges and realise the full potential of the Edge computing paradigm for IoT applications, our community must overcome several critical challenges. The first and foremost is the formulation of decision-making approaches (particularly for resource management and application scheduling) that dynamically determine which subset of data analysis tasks are to be performed on Edge resources and which ones on Cloud data centres to meet the QoS requirements of different IoT applications. Other issues are concerned with how to support multiple tenants with performance isolation between them, to meet the requirements of integrated IoT applications such as smart hospitals.

## 2. Relevant Work

Several frameworks for harnessing Edge and Cloud resources to execute IoT applications have been proposed for many application scenarios, such as smart healthcare [4], smart city [5], smart agriculture [6] and smart grid [7]. They are platform-specific and lack support for the simultaneous execution of multiple applications and offer limited scope for tuning the system according to the specific requirements of applications. To reduce the management overhead, existing frameworks apply centralised techniques that eventually degrade the QoS. In addition, several works discuss different resource provisioning [14][15][19] and application placement policies [11][16][17][18] for Edge computing environments. They provide resources for both latency-sensitive and tolerant applications in similar ways that eventually degrade their performance and are difficult to tune according to the dynamics of real-time IoT applications. Furthermore, contemporary application placement policies for Edge environments are not capable enough to simultaneously deal with the diverse characteristics of different applications meeting their deadline, QoS requirements, user preferences and service provider's interest. While executing applications in a distributed manner, the efficiency of these application placement policies gets affected by node-to-node communication delay, high data sensing frequency of the applications and uneven workloads.

Overall, the existing works lack the seamless management of Edge infrastructures while provisioning resources, placing applications and scheduling IoT workloads to meet the QoS requirements of IoT applications. Many consider Edge and Fog computing paradigms to offer similar capabilities [9] and use them interchangeably. Although consortiums such as OpenFog provide guidelines on the development of such frameworks, they do not offer any algorithms or systems that can simultaneously handle resource provisioning, application placement and workload scheduling in Edge environments. These shortcomings should be addressed by developing new algorithms and systems for handling multiple issues related to QoS, resource provisioning, scheduling, heterogeneity, mobility, faults, and failure management in Edge computing-enabled Smart Hospitals in an integrated manner.

## 3. Significance and Innovation of our Vision

We propose software systems and QoS-based algorithms for provisioning and scheduling real-time IoT applications in Edge computing-enabled Smart Hospitals. We believe this idea is innovative, and original, and highlights significant problems that need to be overcome:

- **Timeliness:** The Internet of Things paradigm is rapidly enabling the creation of smart environments and applications in domains such as healthcare, traffic management, surveillance, and disaster management. However, current Smart Hospital approaches for hosting these applications follow a pure Cloud-centric model (using Data Centre resources located at the endpoint of the network); as a result, they are unable to meet the requirements of real-time decisions due to network latency and congestion issues. These limitations can be overcome by the emerging Edge computing paradigm, which aims at harnessing edge network resources and others as appropriate. Our vision is timely as it aims to develop solutions for the dynamic creation of Edge computing environments and management of Smart Hospital resources and services to meet the QoS requirements of IoT applications. We also must develop mechanisms and policies for a reliable Edge computing environment by ensuring its resilience under failures, performance degradation, and tolerating faults. To achieve this, the management of IoT applications can not just harness edge resources but also Cloud resources for integrated decision-making in contexts such as smart healthcare, where multiple critical infrastructures of IoT applications - such as operating theatre and patient monitoring systems - need to be utilised seamlessly. The central theme of our vision is to develop edge computing technologies for IoT-based healthcare/smart hospital applications in networked and mobile environments. Therefore, we must address issues of relevance to one of the practical research challenges of using technologies effectively as enablers for individuals to manage their healthcare, for instance, using mobile applications, remote monitoring and online access to treatments. While dealing with sensitive hospital data, we should harness the computing capabilities within the edge network to reduce the scope of data exposure to external networks. This will ensure data privacy and add a new dimension to the Cybersecurity area.

- **Originality and Innovation:** Although Edge computing is rapidly growing, software practitioners are facing numerous challenges while creating smart IoT applications that deal with data sources from millions of sensing devices. Currently, little emphasis is paid to the adaptive management of IoT data sources and deployment of IoT applications to harness edge network and Cloud resources seamlessly and reliably and at the same time, meet their QoS requirements. These issues can be addressed by developing new algorithms for integrated management of IoT devices, network edge and end-point resources, and fault-tolerant techniques; hence, its goals are novel and original. In addition, the heterogeneity and topology-aware placement of Smart Hospital application modules can make novel contributions. These advances will enhance the ability of Edge computing to support a wide range of IoT applications.

- **Advancement of the Discipline:** The proposed vision will lead to (1) fundamental principles for the creation of Edge computing environments for QoS-driven IoT applications in the context of Smart Hospitals, (2) novel algorithms for harnessing edge network resources to support latency-sensitive IoT applications, (3) approaches for seamless use of edge and cloud resources for latency-tolerant tasks of integrated IoT applications, and (4) innovative mechanisms and policies for dealing with failures and faults to enhance the resilience of the system along with software technologies. These outcomes will not only advance the Edge computing paradigm but also have an immense impact on distributed systems discipline and its application domains.

- **Significant Business Opportunity:** According to McKinsey's market analysis report "Unlocking the Potential of the Internet of Things", real-time IoT applications will have an economic impact of $11 trillion per year by 2025—equivalent to about 11% of the world economy [3]. A "Health Expenditure Australia 2017–18" report [22] notes that health spending accounted for 10% of overall economic activity. The USA spends 17 percent of its GDP on healthcare each year [21]. Therefore, the realisation of our proposed vision will present a significant business opportunity for companies in the Edge computing and industrial/healthcare IoT applications marketplace.

## 4. Architectural Framework

As stated earlier, this vision paper proposes the creation of QoS-based algorithms for provisioning and scheduling real-time IoT applications in Edge computing-enabled Smart Hospitals. The approach for the realisation of the proposed vision consists of the following steps:

- Architectural framework and principles for Edge computing supporting real-time IoT applications;
- Algorithms for QoS-based provisioning of resources from Edge till the endpoint infrastructure, and techniques for management of performance and failures in virtualised and containerised infrastructures;
- Algorithms for QoS and topology-aware placement and scheduling of IoT applications in Edge computing environments with heterogeneous network and compute resources;
- Mechanisms and policies to increase the resilience of the system;
- Prototype software system and demonstrator IoT-based Smart Hospital applications deployed in Edge computing environments.

*High-Level System Architectural Framework*

The entire solution architecture for managing real-time IoT applications by harnessing both edge network and Cloud resources is shown in Figure 2. The architecture leverages cutting-edge technologies and paradigms to deliver a reliable and scalable Edge computing environment meeting the QoS requirements of IoT applications. We present components of the framework organised at two levels and their related research problems.

1. Gateway Level: This level contains all the components for managing edge resources while deploying the applications, scheduling the IoT workloads, and dealing with the varying application contexts and failures to meet the QoS requirements of applications.
2. Infrastructure Level: This level features computational resources comprising Edge Computational Nodes (ECNs) and Cloud data centres that execute the applications and process the IoT workloads.

*Gateway Level*

IoT Application Broker (IAB) is the core component of the architecture and consists of a set of subcomponents with specialised algorithms. As a mediator between the various medical devices, equipment, or users, and Edge computing network, it is responsible for resource provisioning, scheduling and management of the execution of IoT applications while meeting their QoS requirements. *Data Manager* handles interactions between the gateway node and IoT devices. *QoS-based Resource Provisioner* allocates resources for executing IoT applications based on functional requirements to meet the QoS. *Application Placement Engine* deploys modules of large-scale IoT applications over Edge and Cloud resources. *Monitoring and Resilience Manager* applies policies and techniques to achieve the resilience of the system and deal with failures and performance issues. *IoT Workload Scheduler* schedules IoT workloads for processing in the selected infrastructure. *Directory and Catalogue Services* act as a registry for maintaining instance images and meta-data regarding applications and data flows, and IAB uses its services for resource discovery. The research elements, such as algorithms and mechanisms needed for the realisation of these components, are discussed in the next section.

*Infrastructure Level*

Edge and Cloud Infrastructures will have their own resource managers named Edge Resource Manager (ERM) and Cloud Resource Manager (CRM), respectively. They are responsible for monitoring the context of the respective infrastructure, predicting performance, resource virtualisation (virtual machines and containers), pooling, scaling, coordination and dynamic optimisation (migration and consolidation). In addition, they offer service backup, reliability and fault tolerance during uncertain events such as node failures, resource outages and security attacks. Advances in algorithms and techniques needed for the realisation of ERM are discussed in subsection 5.8, and existing technologies such as OpenStack will serve as CRM.

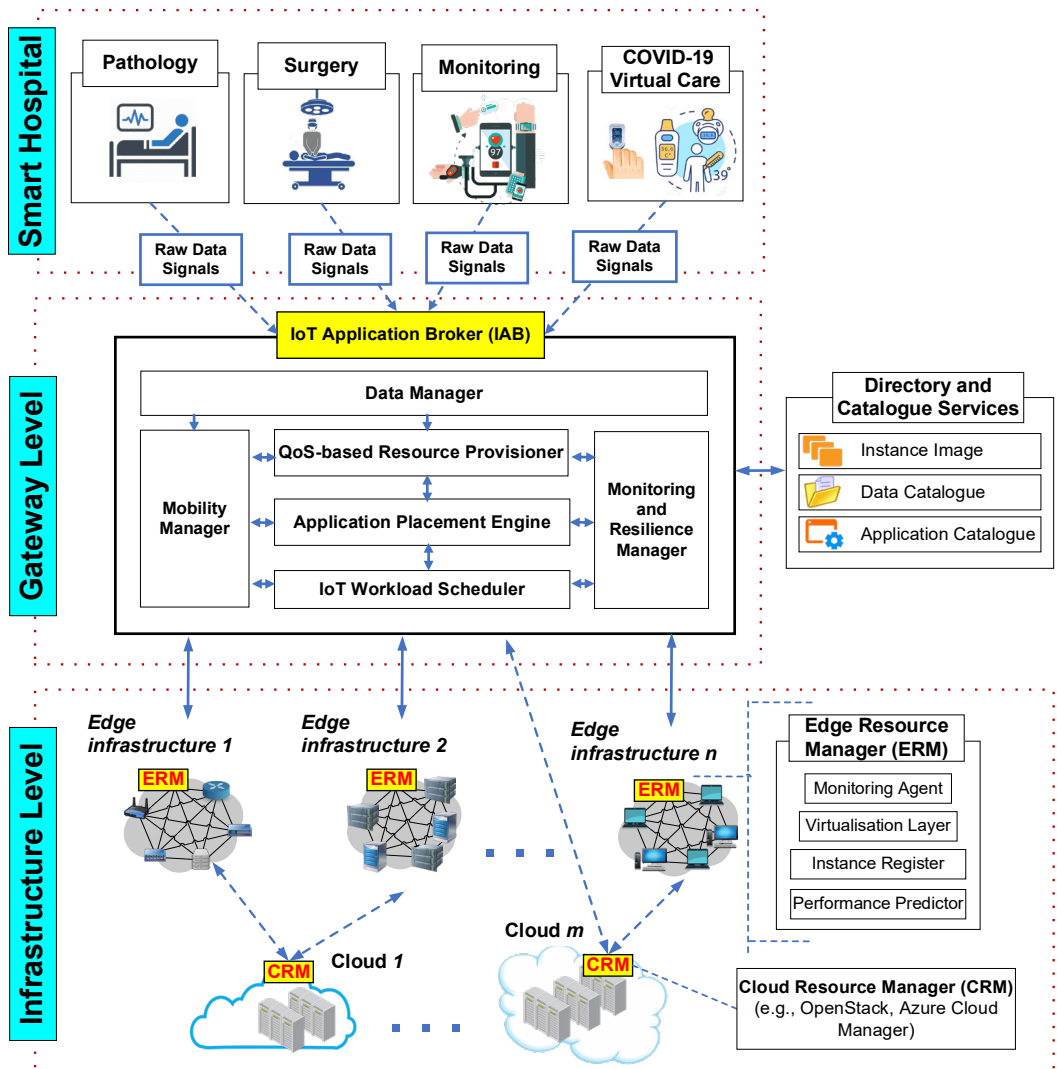

**Figure 2:** Architectural framework for managing IoT applications at Gateway and Infrastructure levels.

## 5. Research Issues and Envisioned Approaches

### 5.1 Data Management Techniques

The basic operations of a Data Manager (DM) include IoT device discovery, sensing frequency calibration, device authentication, data pre-processing and data aggregation from multiple data sources. In Smart Hospitals, the deployment sites and functionalities of IoT devices can be diverse. For example, the pulse oximeter and Electrocardiography system located in the hospital cabins tracks the health status of patients, whereas the surveillance cameras located in the hospital corridors monitor the hospital staff and patient visitors. In the Smart Hospitals context, the greater the amount of sensed data, the higher the accuracy level. During data sensing frequency calibration of IoT devices, both device energy constraints and the required accuracy level should be considered, especially when they are powered by batteries recharged by things like solar panels. Moreover, Smart Hospital workloads are diverse in nature, and tolerable delays for application service delivery can change dynamically and intermittently. In these cases, diverse data streams need to be handled differently to meet the application QoS. Existing data management approaches in Edge computing understate these issues [5]. To address the current limitations, policies should be prepared for making a trade-off between sensing frequency and QoS requirements of the applications so that IoT devices are used in a sustainable/energy-efficient manner. Based on the data flow characteristics, DM scategorises the workloads using historical data analysis. If such categorisation is not effective, DM applies lightweight scategorisation approaches such as workload profiling. DM also deals with patient-centric parameters efficiently so that their dynamics or contextual information (e.g., location, mobility, and criticality) can be effectively utilised during resource provisioning, application placement and workload scheduling.

### *5.2 Algorithms for QoS-based Resource Provisioning*
The QoS-based Resource Provisioner allocates resources in Edge computing environments for executing IoT-based Smart Hospital applications ensuring their guaranteed performance. The QoS requirements of the applications in Smart Hospitals vary from one to another. For example, the tolerable service delivery delay for an application analysing the health parameters of a patient in an emergency unit is more stringent than the application monitoring the health status of outdoor patients. Similarly, the QoS of applications can depend on different types of attributes. For example, when a patient is in motion from the emergency unit to the operation theatre, the real-time and mobility support of the health status analysing application becomes predominant, whereas, during operation, only real-time service delivery drives the QoS of the applications.

Provisioning of resources for applications driven by QoS requirements is a challenging task as there is a waiting time between the moment resources are requested from Edge providers and the time they become available for application execution. This waiting time varies with providers, the number of requested resources, and the load in the computing environment. Existing resource provisioning techniques [10][14][15] did not consider these issues. To overcome this limitation, we forecast the resource requirements and the context variations of QoS-driving attributes of the applications and provide the resources for them in advance. IAB achieves this through negotiation with Edge service providers supporting protocols such as alternate offers [2]. However, depending on various models of the applications, this approach may not be cost-efficient. Alternatively, the provisioner may decide to allocate more powerful resources so that the execution time of the applications gets reduced to meet QoS requirements. Therefore, QoS-based resource provisioning algorithms will dynamically determine the best one from these alternatives. Different providers have different offers in terms of the combination of CPU power, number of cores, memory, and storage capacity of their machine instances. While provisioning, the algorithms should dynamically decide the appropriate combination of resources that meet the QoS requirements of IoT applications.

Furthermore, service providers set different prices for different combinations of resources. To meet user budget constraints, the resource provisioning algorithm should take the combination of resources into account that meets the application QoS requirements at the minimum cost and consumption of energy. Due to the complexity of the problem, multi-criteria optimisation and meta-heuristic algorithms for solving the problem should be created.

### *5.3 Application Placement Algorithms*
Edge computing is constrained in spatial sharing compared to Cloud. For example, private Edge infrastructure is primarily meant for its own use. In such an environment, users' privacy can be better maintained compared to public Clouds, which are vulnerable to security threats/privacy breaches because of the widespread sharing. Since the processing of personal data, such as electronic health data, are subjected to privacy issues, it is expected to perform such operations locally to reduce the scope of privacy breaching. The Application Placement (AP) Engine grasps the privacy preferences of the users and places the applications accordingly over local Edge remote Cloud infrastructure. Moreover, the services of an application can be relevant to many external entities. For example, the health status of a patient can be crucial to insurance companies, employers, specialists and pharmacists. The AP engine ensures ease of access to these application services by placing them in suitable Edge computing nodes. Additionally, Edge computing supports distributed deployment of large-scale applications (such as remote monitoring of the patients within the Smart Hospitals) in a modular way.

Latency sensitivity of different application modules while exchanging data elements through an uncertain network can degrade the service quality. The inter-communication delay among different instances and dependencies between two application modules can play a critical role in this case. Additionally, the sensing frequency of IoT devices can change (e.g. for energy-efficiency of sensors) with time and trigger the execution of other applications that can affect the existing resource management plan significantly. Moreover, the deployed IoT applications may have different priorities, incurring different costs, especially in the healthcare domain. Since all these parameters are tightly coupled, the exploitation of a single parameter while making the application placement decision is not adequate to meet the QoS. This constraint also limits the adaptability of application placement techniques to specific application and network scenarios which is barely dealt with by the existing placement approaches [12] [17]. Our proposed application management algorithms will address these issues by simultaneously considering correlated QoS parameters from the perspective of diverse IoT/Smart Healthcare application scenarios. For the srealisation of this operation, the placement engine will profile the circumstantial significance of different QoS metrics and apply it to dynamically tune the weight of these QoS parameters for varying application service requirements, environmental context, and user demand while dealing with the latency-sensitive applications in Edge computing environments.

*5.4 IoT Workload Scheduling*
When QoS-based provisioning algorithms create a pool of computing instances across multiple Edge infrastructures, application modules get placed in a distributed manner. Once application execution starts over the computing instances, IoT-workload will be scheduled on them. As the Edge environment consists of dynamic and heterogeneous resources, our adaptive scheduling algorithms consider them while mapping tasks/workloads to improve the efficiency of the system. The scheduler also considers data exchange time and network topology while scheduling the workloads. While scheduling workload to multiple applications, IoT Application Broker (IAB) considers their priorities and QoS requirements and makes a rational decision. For example, requests from IoT applications in emergency conditions (such as patients' respiratory problems) will have higher priority than normal use cases (such as sneezing). We schedule workloads across multiple Edge computing instances in a prioritised manner without affecting the service quality requirements of other applications. Our system promotes auto-scaling of the resources while dealing with uneven workloads. In this case, resource provisioning and workload scheduling operations are performed in an integrated manner. As scheduling heterogeneous and uneven workloads can be an NP-complete problem, feasible solutions should be developed to the problem without having any negative impact on users' QoS constraints.

*5.5 Mobility Management*
The mobility patterns of users perceived using embedded GPS of IoT devices or access points' beacons can be a priori-known or a priori-unknown. While managing the mobility of users with a priori-known pattern, the placement engine uses the location trajectories of subscribed patient devices to select the best edge server for the placement and migration of application modules. However, for the priori-unknown mobility scenarios, the placement engine places the applications' modules on the best available server applying predictive analytics on the latest location information of the devices. The mobility manager augments the mobility data to the trajectory history of the users, to identify any possible correlation between the uneven mobility of users and variations within the patient's context. This feature will assist the mobility manager in handling the uneven mobility of the users by exploiting both priori-known and priori-unknown techniques. For each user, it tracks the QoS parameters associated with its application, the available mobility information (such as the history of trajectories and average speed), and the current configuration of application modules assigned to Edge servers through communication with Data Manager, QoS-based Resource Provisioner and Application Placement Engine, respectively. Hence, if critical degradation of QoS parameters occurs, it triggers the application placement engine to find new configurations for application modules using available contextual information of the user and its application and edge devices with sufficient resources.

*5.6 Monitoring and Resilience Management*
To ensure the guaranteed performance of the applications, real-time monitoring of their run-time states is required. A failure of a single application module may compromise the execution of the whole application due to the dependencies. Therefore, we need mechanisms for increasing system resilience, such as module replication and pre-emption. The degree of our module replication depends on constraints noted in QoS parameters such as the budget limit. Such replication of execution will be used to deal with the critical part of the application (e.g., a critical node in a workflow application having a severe impact on overall QoS if a node fails). To support critical applications (e.g. health status monitor during surgery), our system supports pre-emption to prioritise their execution. This is achieved by suspending those applications having minimal impact on their QoS constraints. These techniques, combined with scheduling across multiple Edge infrastructures, make the system resilient in addition to meeting QoS requirements.

Proactive monitoring of resources is critical for ensuring the resilience of the system. Considering the distributed resources in the Edge environment, the likelihood of faults in individual components of the system is high. Moreover, resources are shared among different customers, and this may result in underperformance due to resource contention. Finally, Edge-Cloud resources may not be immediately available, delaying the start of the application. This may render the original plan of replication and pre-emption ineffective to meet QoS expectations and hence, a second line of defence against QoS violations is necessary. Our techniques will monitor and proactively respond to failures and underperformance while ensuring that failures in one or more resources do not compromise the QoS requirements of IoT applications.

*5.7 Directory and Catalogue Services*
Directory and Catalogue services, realisable using Web services technologies such as UDDI (Universal Description, Discovery, and Integration), support publication and discovery of data sources, application services and associated instance images offered by Edge service providers (ESPs). They aid IAB in the discovery of suitable data and application services and their sources during resource provisioning and application scheduling. ESPs allow the creation and storage of computing instance images containing all the software and configurations necessary for the applications. Edge service providers, IAB, and/or users themselves register new images on this catalogue, along with information about software and configurations. Similarly, applications may require access to data that is stored in a repository in a specific location or replicated among a few locations. IAB uses these details maintained in the Data Catalogue to reduce the amount of data movement between the data repository and the applications.

*5.8 Edge Resource Management Policies*
Edge Resource Manager (ERM) is responsible for the management of Edge nodes that are distributed and loosely connected through different networks with varying bandwidth and latency capabilities. Edge nodes can be autonomous or part of a cluster. In the case of dynamic Edge cluster formation, a node hosting ERM monitors contexts such as processing load and energy usage of each node and coordinates their underlying virtualised resources. While provisioning resources and scheduling IoT workload, IAB communicates with the ERM of Edge clusters set up by different Edge Service Providers (ESP). Based on the contextual information, ERM allocates resources for IAB and manages the execution of IoT tasks meeting their QoS requirements.

Since Edge infrastructure is composed of a finite amount of edge resources, excessive resource provisioning requests from IABs can easily surpass the infrastructure capacity. Therefore, application execution experiences delays and this results in poor QoS and SLA violations. Existing policies [4][5] for the management of Edge resources lack a lightweight performance prediction model, platform-independent interactions and admission control mechanisms. ERM will profile the performance of resources while executing different applications and uses platform-independent RESTful APIs for communication. ERM will evaluate the feasibility of applications for execution in Edge based on resource availability and QoS requirements. If a micro-service running on an Edge node requires additional resources and its Edge environment is unable to support it, ERM will delegate latency-sensitive operations to other Edge clusters and latency-tolerant operations to Clouds in collaboration with IAB.

# 6. A Case Study

Our vision leads to the development of algorithms and software systems that support the deployment of IoT applications in Edge computing environments. To demonstrate the feasibility of our vision, we create a prototype system along with an application in a smart hospital scenario, by leveraging existing IoT middleware and associated technologies. Of course, proposed ideas can be first evaluated for various scenarios through modelling and simulation using our iFogSim simulator [13] and then incorporate proven techniques in software systems, including our IoT Application Broker (IAB) and Edge Resource Manager (ERM). To reduce the cost of software engineering of IAB and ERM, we will leverage existing IoT middleware and associated technologies. They include (a) ZigBee and Constrained Application Protocol (CoAP) for communication between IoT sensors and Gateway, and Web services-based interfaces for interaction between components in Gateway and Infrastructure levels, (b) OpenHAB, Apache Edgent, and others such as Fogbus [20], (c) Docker containers for creating micro-services and its orchestration engine for management of Edge nodes, and (d) Cloud resource management technologies such as OpenStack.

*6.1 A Smart Healthcare Application and Edge Computing Environment*
We will now discuss a case study of creating a sample edge computing environment using the Fogbus2 framework [23], and an application demonstrator in the context of diabetes monitoring and prediction. FogBus is a framework for scheduling and processing heterogeneous IoT applications in a distributed manner. The network topology in FogBus is based on a master-worker model in which the master controls the created task executor nodes (i.e., workers). It connects different IoT devices and sensors with edge and cloud infrastructure via a gateway device, to process data and tasks on worker nodes in the edge and cloud environment. Figure 3 represents our case study for diabetes monitoring and prediction using FogBus. In this IoT application, IoT devices for health monitoring, known as the Internet of Medical Things (IoMT), are used to sense diabetes risk factors, such as blood pressure, and send the data over the internet to an Edge computing node for pre-processing and predicting the prevalence of diabetes for a user. Streamed data is sent by the edge computing node to the cloud node for storage. The prediction model on the edge computing node is the result of machine learning training in the Cloud. Training the diabetes dataset in the Cloud is achieved offline so that the development of the prediction model does not impact the prediction performance on the Edge. The prediction model on the Edge is updated whenever a new model is produced by the Cloud. The diabetes prediction IoT application uses Fogbus, which consists of various hardware and software components as described below:

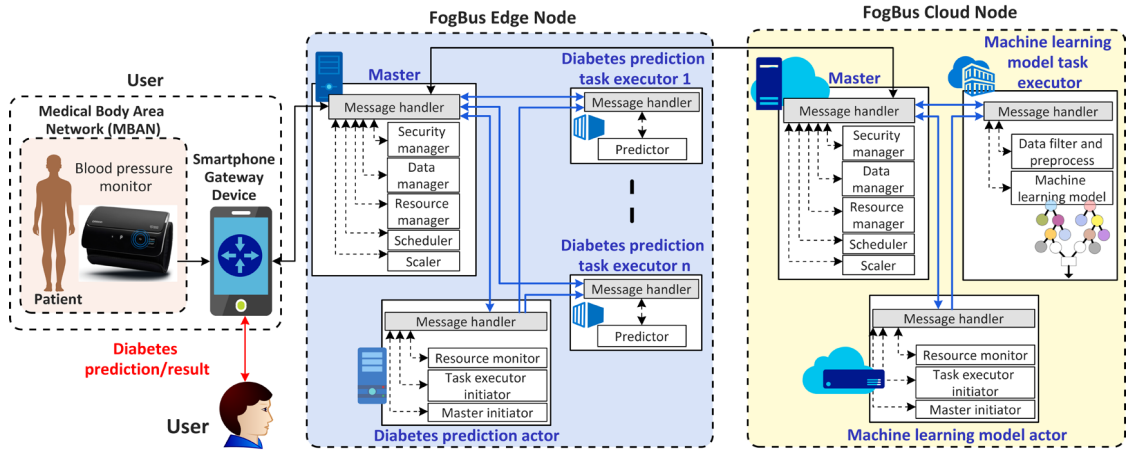

**Figure 3:** Smart Healthcare Diabetes monitoring and prediction Architecture.

Hardware components

- Medical Body Area Network (MBAN): It consists of multiple IoMT sensor nodes, where each node collects, samples, and communicates biomedical information. These nodes are low-powered IoMT devices with limited computing and storage capabilities. The collected information is transmitted to the connected gateway devices using different communication protocols such as WiFi, Zigbee, and Bluetooth. Different IoMT devices are used to measure different diabetes risk factors, such as hypertension, obesity, cholesterol level, depression, serum uric acid level, physical activity, and glucose level. In this case study, we consider a hypertension monitor, Omron EVOLV HEM-7600T-E, that measures the systolic and diastolic blood pressure of a patient. The measurements are then sent to a smartphone gateway device.
- Gateway Device: The smartphone gateway device receives the hypertension data transmitted by the medical sensors. This information includes a timestamp, systolic blood pressure, and diastolic blood pressure. The data is then forwarded to the edge node for further processing.
- Edge Node: FogBus provides the gateway devices with low-latency and high-bandwidth access to one or multi-hop away Edge node, consisting of heterogenous edge servers, for real-time processing and diabetes prediction. Edge resources close to the IoMT devices support the stringent requirements of smart healthcare applications in terms of latency and throughput. In our case study, the edge node uses a trained machine learning model to predict the prevalence of diabetes in a user based on the values of risk factors.
- Cloud Node: FogBus extends the computing and storage capabilities of Edge by providing access to multiple Cloud nodes. Cloud provides scalable resources in terms of processing and storage, supporting computing requirements of IoMT applications such as data analytics for prognosis/diagnosis of chronic diseases. In our study, cloud servers are for data storage and big data analytics. They train machine learning models using the data collected from sensors. The trained model is sent back to the Edge for real-time prediction. Our case study uses the Random Forest machine learning model as it is found to be the most accurate for diabetes prediction [24].

Software components

- Master: This component runs on both edge and cloud nodes. The master component at the edge node receives the following information 1) the input hypertension data from the gateway devices, 2) the trained machine learning model for prediction from the master component at the cloud node, and 3) the diabetes prediction requests from the users. The master at the cloud node receives the data from the edge node to train a machine-learning model for diabetes prediction. The security manager module within the master aids in secure communication between different components and ensures sauthorised data access. The data manager module communicates the data, prediction model and prediction results with the corresponding component. The resource manager module receives information regarding the available computing resources of all task executors (i.e., CPU utilisation, memory utilisation, and disk i/o), network characteristics (i.e., latency, network i/o, and bandwidth), and requirements of IoMT applications (i.e., computing power and deadline). The scheduler receives the list of actors and schedules the task request, i.e., machine learning model development or diabetes prediction, to one of the available actors. The scaler module is used to initiate a new master for scalability.

- Actor: This component runs on edge and cloud nodes. The task executor initiator module is called whenever the master component assigns a model development task or a diabetes prediction task to an actor on the edge and cloud nodes, respectively. It initiates a task executor component and defines where the result (i.e., prediction result or prediction model) must be forwarded by the task executor. The master initiator component is called when the scaler module of the master component is executed to scale up the application's resources. It initiates a new master component on a new host and receives the list of actors from the master component on which the scaler module had executed. The resource monitor module collects the resource sutilisation values of associated task executors in real-time and sends them to the resource manager module within the master. In our case study, there are two actor components: 1) the diabetes prediction actor which runs on Edge nodes and is responsible for initiating a task executor for predicting the prevalence of diabetes using a trained machine learning model, and 2) the Machine Learning model actor: This component runs on cloud nodes and is responsible for initiating a task executor to develop a machine learning prediction model.
- Task executor: This component performs the tasks allocated by the actor component. The diabetes prediction task executor involves a predictor module which uses the developed prediction model to predict the prevalence of diabetes for a user based on the input data from hypertension monitoring IoT device, and the machine learning model task executor is a machine learning model module which trains a diabetes dataset and develops the diabetes prediction model based. The data filtering and pre-processing module filters the data by removing irrelevant observations to diabetes machine learning development.

### 6.2 A Deployment Application Demonstrator

We demonstrate our smart healthcare diabetes monitoring and prediction IoMT applications. As shown in Figure 4, the deployment consists of a blood pressure IoMT device, a diabetes prevalence risk predictor on the Edge, and a machine learning prediction model builder in the Cloud. We use Omron EVOLV HEM-7600T-E blood pressure monitor that measures the systolic and diastolic pressure of the user. It senses and sends the blood pressure data to the smartphone gateway device using Bluetooth communication. In turn, the smartphone filters the data by removing the user's identity for privacy. The timestamp, systolic blood pressure, and diastolic blood pressure are displayed on the gateway device. The anonymous data is forwarded to the edge node for predicting the risk of diabetes prevalence. The prediction model is trained offline in the Cloud.

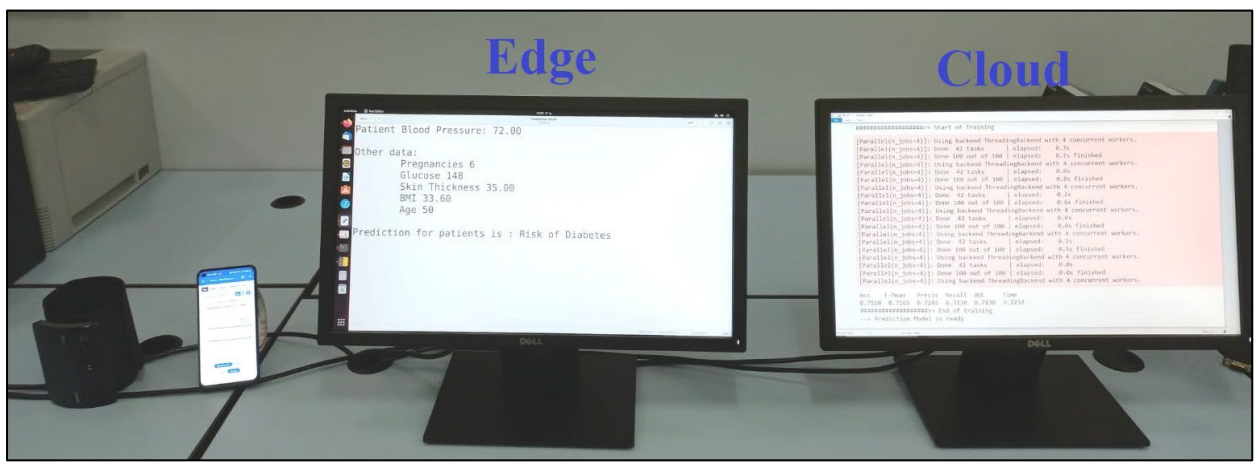

**Figure 4:** Smart Healthcare Diabetes monitoring and prediction deployment.

**Diabetes Dataset**

We consider the Pima Indian Diabetes dataset from the National Institute of Diabetes and Digestive and Kidney Diseases. The dataset diagnostically determines whether a patient is diabetic or non-diabetic based on medical measurements. The dataset includes information for female patients of Pima Indian Heritage who are at least 21 years old. It consists of 9 features as follows. (1) pregnancies: number of times pregnant, (2) glucose: plasma glucose concentration at 2 hours in an oral Glucose Tolerance Test (GTT), (3) diastolic blood pressure (mm Hg),

(4) triceps skin fold thickness (mm), (5) 2-hour serum insulin (µU/ml), (6) Body Mass Index (BMI) (weight in kg/(height in m)$^2$), (7) diabetes pedigree function: expected genetic influence of diabetic and non-diabetic relatives (parents, grandparents, full siblings, half-siblings, full aunts and uncles, half aunts and uncles, and first cousins) on the patient's eventual diabetes risk, (8) age: age (years), and (9) outcome: diagnosis of diabetes – value 1: diabetic and value 0: non-diabetic. The dataset consists of medical records of 768 patients, out of which 268 (34.9%) are diabetic and 500 (65.1%) are non-diabetic. Figure 5 shows the distribution of Numerical features for the diabetic and non-diabetic classes [24]. Table 1 shows a sample of the Pima Indian dataset for 10 patients.

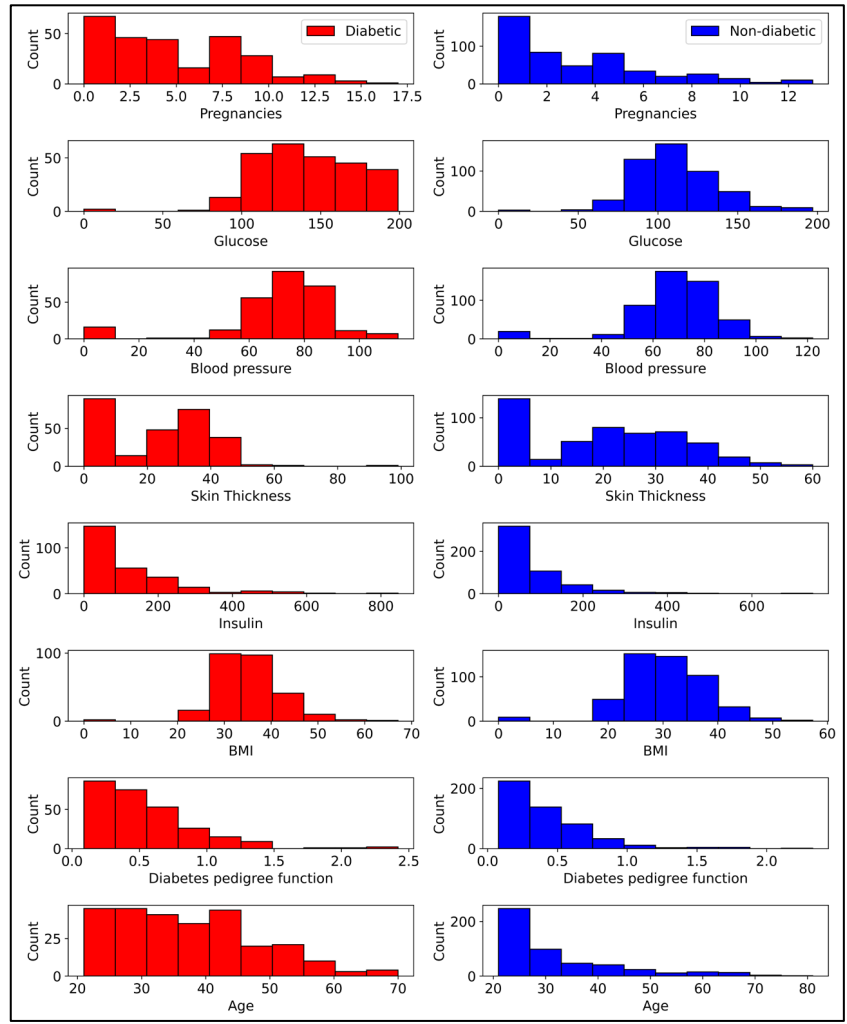

**Figure 5:** Distribution of numerical features for diabetic and non-diabetic classes.

**Table 1:** Sample patient record data from the Pima Indian dataset.

| Pregnancies | Glucose | Diastolic blood pressure | Triceps skin fold thickness | Insulin | Body Mass Index | Diabetes pedigree function | Age | Outcome |
|---|---|---|---|---|---|---|---|---|
| 1 | 89 | 66 | 23 | 94 | 28.1 | 0.167 | 21 | 0 |
| 0 | 137 | 40 | 35 | 168 | 43.1 | 2.288 | 33 | 1 |
| 3 | 78 | 50 | 32 | 88 | 31 | 0.248 | 26 | 1 |
| 2 | 197 | 70 | 45 | 543 | 30.5 | 0.158 | 53 | 1 |
| 1 | 189 | 60 | 23 | 846 | 30.1 | 0.398 | 59 | 1 |
| 5 | 166 | 72 | 19 | 175 | 25.8 | 0.587 | 51 | 1 |
| 0 | 118 | 84 | 47 | 230 | 45.8 | 0.551 | 31 | 1 |
| 1 | 103 | 30 | 38 | 83 | 43.3 | 0.183 | 33 | 0 |
| 1 | 115 | 70 | 30 | 96 | 34.6 | 0.529 | 32 | 1 |
| 3 | 126 | 88 | 41 | 235 | 39.3 | 0.704 | 27 | 0 |

The dataset is pre-processed by removing observations that have missing values. As shown in Figure 5, there exist observations with a '0' value for skin thickness, blood pressure, and BMI features, indicating missing values. Consequently, the pre-processed dataset includes 537 records with 179 (33.3%) diabetics and 358 (66.7%) non-diabetics. The pre-processed file is saved in Comma Separated Value (.csv) file.

**Machine learning Model: Training and Deployment**
We use the Random Forest machine learning model for diabetes prediction as it outperforms other machine learning models for diabetes prediction [25], using the diabetes dataset in .csv format, as shown in Figure 6. The model is trained in the Cloud and the prediction model is deployed on the Edge.

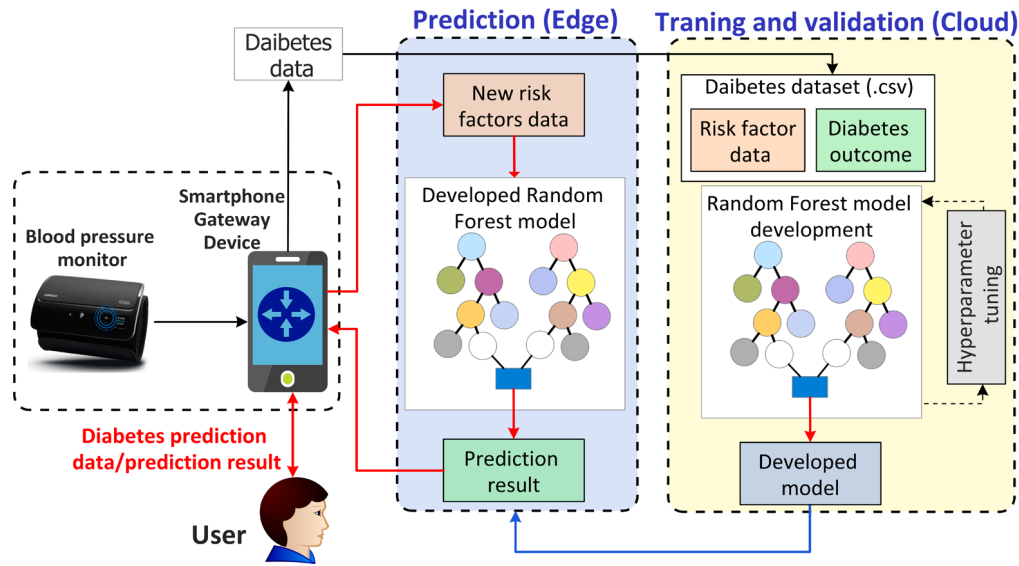

**Figure 6:** Machine learning model development and prediction.

For model development and validation, we divide the PIMA Indian dataset into 70% (training dataset) and 30% (validation dataset) respectively. The training dataset is used to develop a diabetes prediction model and the validation dataset is used to evaluate the performance of the developed model. We implement the Random Forest machine learning model using Scikit learn library of Python 3.10 programming language with the following optimal parameters, 100 n_estimators (the number of decision trees in the forest), 5 max_depth (the maximum depth of each tree), all features max_features (the number of features to be considered when looking for the best split), and entropy criterion (function to measure the quality of a split).

## 7. Conclusions and Final Remarks

The rising cost of healthcare has been a major concern in most countries. For the past decade, the United States has spent more than 17 percent of its GDP on healthcare each year [21]. A recent Australian Institute of Health and Welfare report [22] noted that health spending accounted for 10% of overall economic activity in Australia during 2017-18 and it increased dramatically due to the COVID-19 pandemic. As this vision paper proposes new Edge computing technologies for enabling IoT applications that occur in areas such as Smart Hospitals and Healthcare, it presents significant national benefits by increasing the efficiency of hospital/healthcare management. The central theme of the proposed vision is to enable the creation of technologies for reliable and QoS-driven execution of IoT applications in networked and mobile environments for healthcare/smart hospitals applications.

Hospitals' existing reliance on cloud computing has certainly begun to transform this sector, but at the same time has intorcuded certain challenges, which our vision begins to address. For example, hospitals are increasingly dealing with the phenomenon of staff bringing their personal devices to hospitals, and using them to create and share sensitive information. One way to overcome this challenge is to use microservices on staff's personal devices, which can be used to capture and transmit data safely. Another challenge that hospitals face is transitioning from in-ward to in-home care. In such cases where patients can return home and keep being treated, perhaps they could be given an edge device that can act as a gateway to collecting and transmitting sensitive data to the hospital cloud while also allowing for microservice placement policies to be implemented.

The problem addressed by this vision benefits mission-critical applications in several other domains, including smart cities, smart transport, and smart agriculture. As IoT applications are projected to have an economic impact of $11 trillion per year by 2025 (equivalent to about 11% of the world economy), the realisation of the proposed

vision will create a significant business opportunity for businesses in computing technologies and the IoT applications marketplace.

This vision paper's objective is to propose approaches for the creation of algorithms and software systems for managing Edge computing resources and efficient deployment of IoT applications in Smart Hospital's domains to meet their QoS requirements. To achieve this objective, our future directions entail the following guidelines:

1. Create an architectural framework and principles for Edge computing for IoT applications in Smart Hospitals.
2. Propose new algorithms for QoS-based provisioning of resources in Edge computing environments for IoT-based Smart Hospital applications and techniques to monitor and manage virtualised infrastructures.
3. Propose new algorithms for scheduling IoT applications in Edge computing environments by considering diverse topology and contextual variations such as doctor mobility, criticality, and privacy issues in Smart Hospitals.
4. Develop mechanisms and policies for task pre-emption, resource replication, and monitoring and reactive approaches to enhance system resilience under failures or performance issues.
5. Develop a software platform – incorporating the above mechanisms and techniques – and deploy it within the Melbourne Smart Hospital Living Lab and provide real-world demonstrator real-time IoT applications.

In summary, the realisation of this proposed vision will lead to (a) architectural principles for Edge computing for real-time IoT applications, (b) innovative algorithms for scheduling IoT applications on the Edge and Cloud resources, (c) novel software technology for the management of Edge resources, and (d) application demonstrators in healthcare/Smart hospitals area.

**References**
[1] J. Gubbi, R. Buyya, S. Marusic, and M. Palaniswami. *Internet of Things (IoT): A Vision, Architectural Elements, and Future Directions*. Future Generation Computer Systems, 29(7):1645-1660, September 2013.
[2] S. Venugopal, X. Chu, and R. Buyya. *A Negotiation Mechanism for Advance Resource Reservation using the Alternate Offers Protocol*. Proceedings of the 16th International Workshop on Quality of Service, Twente, The Netherlands, June 2008.
[3] J. Manyika, M. Chui, P. Bisson, J, Woetzel, R. Dobbs, J. Bughin, and D. Aharon. *Unlocking the Potential of the Internet of Things*. McKinsey & Company, 2015.
[4] A. Rahmani, T. Gia, B. Negash, A. Anzanpour, I. Azimi, M.Jiang, and Pasi Liljeberg. *Exploiting smart e-Health gateways at the Edge of healthcare Internet-of-Things: A fog computing approach*. Future Generation Computer Systems, 78:641–658, 2018.
[5] D. Bruneo, S. Distefano, F. Longo, G. Merlino, A. Puliafito, V. D'Amico, M. Sapienza, and G. Torrisi. *Stack4Things as a Fog computing platform for Smart City applications*. Proceedings of the IEEE Conference on Computer Communications Workshops, San Francisco, CA, USA, 2016
[6] M. Roopaei, P. Rad, and K. Choo. *Cloud of Things in Smart Agriculture: Intelligent Irrigation Monitoring by Thermal Imaging*. IEEE Cloud Computing, 4(1): 10-15, IEEE Press, USA, 2017.
[7] S. Bera, S. Misra and J. Rodrigues. *Cloud Computing Applications for Smart Grid: A Survey*. IEEE Transactions on Parallel and Distributed Systems, 26(5): 1477-1494, May 2015.
[8] L. Belli, S. Cirani, G. Ferrari, L. Melegari, and M. Picone. *A Graph-based Cloud Architecture for Big Stream Real-time Applications in the Internet of Things*. Proceedings of the European Conference on Service-Oriented and Cloud Computing, Manchester, UK, 2014.
[9] A. Zomaya. *Resource Management in Edge Computing: Opportunities and Open Issues*. Proceedings of the 2019 IEEE Symposium on Computers and Communications (ISCC), Barcelona, Spain, July 2019.
[10] M. Aazam and E. Huh. *Dynamic Resource Provisioning through Fog Micro Datacentre*. Proceedings of the IEEE International Conference on Pervasive Computing and Communication Workshops. St. Louis, USA, March 2015.
[11] C. Li, J. Tang, H. Tang, and Y. Luo. *Collaborative cache allocation and task scheduling for data-intensive applications in edge computing environment.* Future Generation Computer Systems, 95:249-264, June 2019.
[12] F. Juan, and A. Ma. *IoT Application Modules Placement and Dynamic Task Processing in Edge-Cloud Computing*. IEEE Internet of Things Journal, 8(16): 12771-12781, August 2021.
[13] H. Gupta, A. Dastjerdi, S. Ghosh, and R. Buyya. *iFogSim: A Toolkit for Modeling and Simulation of Resource Management Techniques in Internet of Things, Edge and Fog Computing Environments*. Software: Practice and Experience, 47(9):1275-1296, 2017.
[14] S. Dai, L. Hai, Y. Li, and Z. Zhang. *An Incentive Auction-based Cooperative Resource Provisioning Scheme for Edge Computing over Passive Optical Networks*. Proceedings of the 18th International Conference on Optical Communications and Networks (ICOCN), 2019.

[15] C. Avasalcai and S. Dustdar. *Latency-aware distributed resource provisioning for deploying iot applications at the Edge of the network*. Proceedings of the Future of Information and Communication Conference, pp. 377-391. Springer, Cham, 2019.

[16] R. Mahmud, S. Srirama, K. Ramamohanarao, and R. Buyya. *Profit-aware Application Placement for Integrated Fog-Cloud Computing Environments*. Journal of Parallel and Distributed Computing (JPDC), 135: 177-190, January 2020.

[17] C. Li, J. Bai, and J. Tang. *Joint soptimisation of data placement and scheduling for improving user experience in edge computing*. Journal of Parallel and Distributed Computing, 125: 93-105, March 2019

[18] M. Goudarzi, H. Wu, M. Palaniswami, and R. Buyya. *An Application Placement Technique for Concurrent IoT Applications in Edge and Fog Computing Environments*. IEEE Transactions on Mobile Computing, 20(4): 1298-1311, April 2021.

[19] T. Bahreini, H. Badri, and D. Grosu. *Energy-aware Capacity Provisioning and Resource Allocation in Edge Computing Systems*. Proceedings of the International Conference on Edge Computing, 2019.

[20] S. Tuli, R. Mahmud, S. Tuli, and R. Buyya. *FogBus: A Blockchain-based Lightweight Framework for Edge and Fog Computing*. Journal of Systems and Software, 154: 22-36, August 2019.

[21] B. Chen, A. Baur, M. Stepniak, and J. Wang. *Finding the future of care provision: The role of smart hospitals*, MCKinsey Report, May 2019.

[22] AIHW. *Health Expenditure Australia 2017–18*, ISSN: 2205-6610, Canberra, Sep. 25, 2019.

[23] Q. Deng, M. Goudarzi, and R. Buyya. *FogBus2: A Lightweight and Distributed Container-based Framework for Integration of IoT-enabled Systems with Edge and Cloud Computing*. Proceedings of the SIGMOD 2021 International Workshop on Big Data in Emergent Distributed Environments, June 20-25, 2021.

[24] A. Hennebelle, H. Materwala, and L. Ismail. *HealthEdge: A Machine Learning-Based Smart Healthcare Framework for Prediction of Type 2 Diabetes in an Integrated IoT, Edge, and Cloud Computing System*. Proceedings of the 14th International Conference on Ambient Systems, Networks and Technologies, 2023.

[25] L. Ismail, H. Materwala, M. Tayefi, P. Ngo, and A. P. Karduck. *Type 2 Diabetes with Artificial Intelligence Machine Learning: Methods and Evaluation*. Archives of Computational Methods in Engineering, 29(1): 313–333, Jan. 2022.